\documentclass[final,3p,twocolumn,authoryear]{elsarticle}
\usepackage{graphicx}
\usepackage{amssymb}
\usepackage{amsmath}
\usepackage{xcolor}
\usepackage[unicode,pdfhighlight=/P,colorlinks=true]{hyperref}

\newcommand{\Geo}[2]{\ensuremath{#1^\circ\,\text{#2}}}
\newcommand{\E}[1]{\Geo{#1}{E}}

\newcommand{\N}[1]{\Geo{#1}{N}}

\begin{document}
\hypersetup{colorlinks=true,linkcolor=red,citecolor=brown,urlcolor=blue,unicode,pdfhighlight=/P}
\begin{frontmatter}


\title{Lagrangian analysis of the vertical structure of eddies \\
simulated in the Japan Basin of the Japan/East Sea}


\author{S.V. Prants}
\ead{prants@poi.dvo.ru}

\author{V.I. Ponomarev}
\author{M.V. Budyansky}
\author{M.Yu. Uleysky}
\author{P.A. Fyman}

\address{Pacific Oceanological Institute of the Russian Academy of Sciences,\\
43 Baltiiskaya st., 690041 Vladivostok, Russia\\
URL: \url{http://dynalab.poi.dvo.ru}}

\begin{abstract}
The output from an eddy-resolved multi-layered
circulation model is used to analyze
the vertical structure of simulated deep-sea eddies
in the Japan Basin of the Japan/East Sea constrained by bottom topography.
We focus on Lagrangian analysis of anticyclonic eddies,
generated in the model
in a typical year approximately at the place of the M3 mooring \citep{Takematsu99}
and the hydrographic sections \citep{Talley2001}, where such eddies have been
regularly observed in different years (1993--1997, 1999--2001).
Using a quasi-3D computation of
the finite-time Lyapunov exponents and displacements for
a large number of synthetic tracers in each depth layer,
we demonstrate how the simulated feature evolves of the eddy, that does not reach the surface
in summer, into a one reaching the surface in fall.
This finding is confirmed by
computing deformation of the model layers across the simulated eddy
in zonal and meridional directions and in the corresponding temperature cross sections.
Computed Lagrangian tracking maps allow to trace the origin
and fate of water masses in different layers of the eddy.
The results  of simulation are compared with observed temperature zonal and meridional
cross sections of a real anticyclonic eddy to be studied at that place
during the oceanographic Conductivity, Temperature, and Depth (CTD) 
hydrochemical survey in summer 1999
\citep{Talley2001}. Both the simulated and observed eddies are shown to have the similar
eddy core and the relief of layer interfaces and isotherms.
\end{abstract}

\begin{keyword}
3D Lagrangian analysis \sep numerical circulation model of the Japan/East Sea \sep vertical structure of eddies


\end{keyword}
\end{frontmatter}

\section{Introduction}

\begin{figure*}[!htb]
\begin{center}
\includegraphics[width=0.8\textwidth,clip]{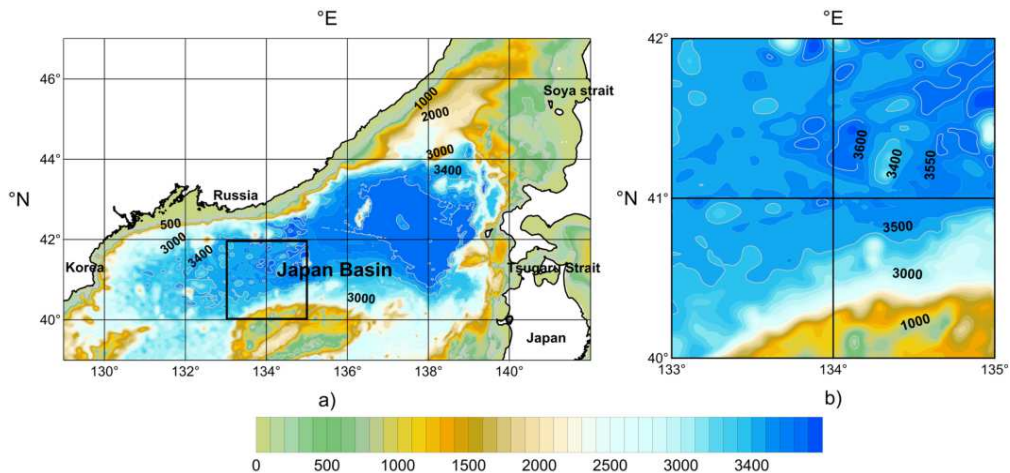}
\end{center}
\caption{The bottom topography in a) the northern part of the Japan/East Sea
and b) the studied area.}
\label{fig1}
\end{figure*}

The Japan/East Sea (JES) is the semi-closed marginal sea with narrow shallow
straits and three deep basins. The Proper Water of the JES with a
weak stratification is formed from the surface waters basically during
winter in the deepest Japan Basin (JB, 3000--3560~m). It
occupies the northern part of the sea (Fig.~\ref{fig1}) where the climate is more severe
than in the southern one. The general circulation pattern of the JES,
to be found from observations \citep[see, e.~g.,][and references therein]{Takematsu99,Lee2005} 
and modeling \citep[see, e.~g.,][and references therein]{Holloway1995,Yoon2009},
includes the cyclonic gyre with several cores of the large-scale cyclonic
patterns over the JB \citep[see, e.~g.,][and references therein]{Trusenkova2005} and a current
system with prevailing anticyclonic vorticity in the southern sea area
\citep[see, e.~g.,][and references therein]{Lee2005,Lee2010}.
\begin{figure*}[!htb]
\begin{center}
\includegraphics[width=0.7\textwidth,clip]{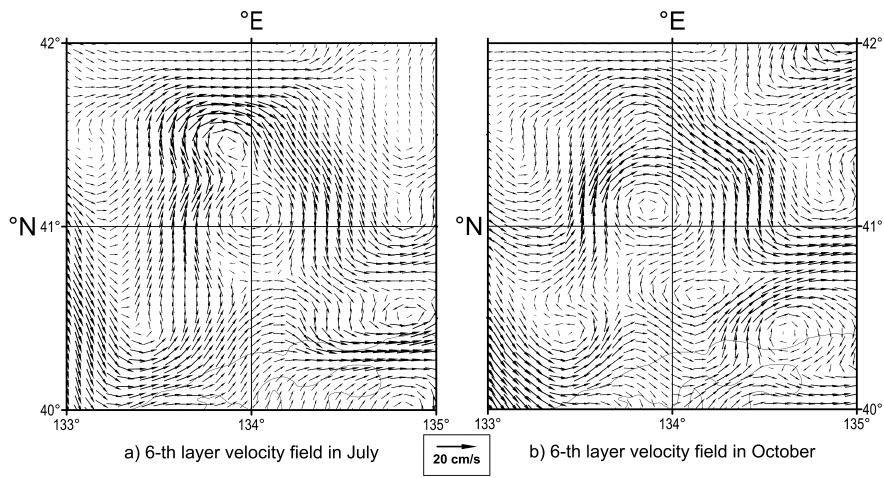}
\caption{Simulated current velocity field in the studied area
in the 6th layer velocity field averaged over a) July and b) October.}
\label{fig2}
\end{center}
\end{figure*}

The Subpolar Front (SPF) separates subtropical and subarctical waters
and lies between \N{38} and \N{41}
being a boundary of distinct physical and chemical properties and a distinct vertical
water structure \citep{Talley2006}.
The zonal eastward current along the SPF is known to be highly variable with
seasonal migration \citep{Yoon2009}. Meandering of the SPF is accompanied
with formation of eddies of different polarity and sizes
generated through baroclinic instability \citep{Lee2005,Lee2010}.
Satellite infrared images and drifter observations show that
the eastward extension of the East Korean Warm Current along the SPF
looks like a typical frontal eddy street (the scale is about 100~km)
over the southern edge of the JB. The mesoscale and submesoscale eddies
with smaller scales 30--50~km and 3--10~km are formed, respectively, over
the JB and its steep northwestern continental slope and shelf break \citep{OM11,FAO13}.


Anticyclonic eddies have been often observed to the north of the SPF.
As part of an international cooperative program Circulation Research of 
the East Asian Marginal Seas (CREAMS), long-term moored current measurements have been
carried out at seven sites in the JB. One of the moorings,
M3, was deployed at \N{41.5} and \E{134.3} where the JB is 3500~m deep.
It was equipped with three current meters at about 1000, 2000 and 3000~m depths
which made the measurements for three years, from August 1993 to July 1996,
with the data sampling period of one hour.
The current meter data of 3-year duration have shown  that deep-sea
anticyclonic eddies (ACEs) with the orbital speeds of the order of 0.1~m~s$^{-1}$ occurred every year
in the deep layers \citep{Takematsu99}. Available time
series of sea-surface-temperature images and World Ocean Circulation Experiment 
(WOCE) drifter tracks were well
correlated to that finding. The currents at 1000,
2000, 3000~m have been found by \citep{Takematsu99} to be highly coherent throughout
the observational period of August 1993 to July 1996.
They have observed intensification of the current in fall and winter.
The eddies observed in the JB did not exhibit any definite direction of propagation.
\citep{Takematsu99} indicated that the effect of the bottom geometry may
be important. Once a barotropic eddy is established in
the JB, it will be forced to remain only in the
interior region of the Basin. In fact, the eddy currents were
observed only at M3, but not in the rim area of the Basin
at M1, M2, M4, M6 and M7 stations \citep{Takematsu99}.

During the oceanographic CTD-hydrochemical survey
in summer 1999 \citep{Talley2001,Talley2006}, the mesoscale ACE with the center
approximately at the site of the M3 mooring has been found in
temperature, salinity, dissolved oxygen and NO${}_3$ sections 
along \E{134} and \N{41.25}.

In this paper we use the output from the eddy-resolved multi-layered
circulation MHI model \citep{Shapiro2000} to analyze from a Lagrangian
perspective the vertical structure of simulated deep-sea ACEs in the JB constrained by bottom topography.
We focus on the ACE, generated in the MHI model approximately
at the place of the M3 mooring \citep{Takematsu99} and the hydrographic
sections \citep{Talley2001,Talley2006}, where such eddies have been regularly
observed in different years (1993--1997, 1999--2001).

Two-dimensional Lagrangian approach has been successfully applied for
studying horizontal transport and mixing in numerically-generated
and altimetric velocity fields in different basins, from bays \citep{LC05,LS06,GF09,FAO13}
and seas \citep{AR02,OF04,Schneider2005,MH08,Prants2013}
to the ocean scale \citep{BO08,Waugh2008,P13,Prants2014}.
Lagrangian methods are especially suitable
because they allow to delineate eddy's boundaries \citep{OM11,Haller13,Prants2014},
to visualize transport barriers and corridors
along which the core of an eddy is exchanged by water with its surrounding
\citep{Kirwan2013,FAO13,Prants2013} and to quantify that transport \citep{Miller}.

The challenging problem is identifying the
vertical structure of an eddy and quantifying its coherence.
In order to quantify properties of the 3D structure of eddies (for example, their volume),
it is often the eddy's surface edges are simply extended to a given depth along the vertical.
It is well known, however, that most eddies do not have a cylinder-like form. Moreover,
the intriguing
problem is changes in the structure of eddies at different depths in the course of time.

The Lagrangian analysis in this paper will be performed using two indicators,
the finite-time Lyapunov exponent (FTLE) \citep{Pierrehumbert} and displacements for
a large number of tracers \citep{DAN11,P13,Prants2014}.
We will show how our modeled ACE evolves of the eddy without any signs of
rotation motion at the sea surface in summer into a one reaching the surface in fall.
In order to demonstrate that we implement a quasi-3D computation of those Lagrangian indicators.
We use the full 3D velocity field from the output of the MHI model and compute the synoptic maps
of the FTLE and particle's displacements in every model layer.

Quasi-3D Lagrangian approach has been applied recently by \citep{Bettencourt12}
for diagnostics of 3D Lagrangian coherent structures around a particular cyclonic eddy
pinched off from a Benguela upwelling front.
Three-dimensional Lagrangian coherent structures were extracted
as ``ridges'' of the calculated
fields of the finite-size Lyapunov exponent obtained from an output of 
the Regional Ocean
Modeling System (ROMS). They have been found by \citep{Bettencourt12} to be quasi-vertical surfaces.
Another eddy feature, a ring of the Loop Current in the Gulf of Mexico has been studied
by \citep{Kirwan2013} by the
similar method, using the data-assimilating
HYbrid Coordinate Ocean Model (HYCOM). Those authors have studied the location of relevant
transport barriers during the formation of Eddy Franklin in 2010 at several depths from the
surface down to 200~m.


Our paper is organized as follows. Section~2 describes briefly the eddy-resolved multi-layered
circulation MHI model adopted to the JES and the Lagrangian approach to be used.
Section~3 contains the main results of the 3D Lagrangian analysis
of isolated deep-sea eddies in the JB.
As the study case, we focus on a simulated mesoscale ACE to the north of the SPF
and investigate its vertical structure.
Backward-in-time FTLE and drift maps are calculated in different layers in Section~3.1
to illustrate the evolution of boundaries of the eddy and its vertical structure
during summer and fall. The deformation of the model layers across the eddy
in zonal and meridional directions and the corresponding temperature cross sections
are computed in Section~3.2 and compared in Section~3.3 with the data of the oceanographic
CTD-hydrochemical survey in summer 1999 \citep{Talley2001,Talley2006}.
In Section~3.4 we calculate the Lagrangian tracking
maps which allow to illustrate pathways and barriers to transport into and out
of the eddy at different depths. The form, nonlinearity and stability of simulated
ACEs are discussed briefly in the end of Section~3.

\section{The model and methods}

\subsection{Circulation numerical model adopted to the Japan/East Sea}

We use the MHI ocean circulation quasi-isopicnal layered model
\citep{Mikhailova1993,Shapiro2000} with a free surface boundary condition
incorporating the horizontally inhomogeneous upper mixed layer.
The model is based on a system of primitive equations integrated within
each quasi-isopicnal layer. All layers are assumed to be horizontally inhomogeneous,
however, the density in each thermocline layer changes within the limits determined
{\em a priori} by the prescribed basic stratification. 

It is assumed that the layers may outcrop. The layer outcropping is similar to the 
isopycnal model applied by \citep{Hogan2006} to the Japan/East Sea. 
The interfaces of the inner model layers can climb to the upper mixed layer in the 
frontal zones. The horizontally inhomogeneous upper-mixed-layer model 
includes parametrizations of turbulent heat, salt and momentum fluxes, drift current 
in that layer, entrainment and subduction processes at the bottom of the layer 
\citep{Mikhailova1993,Shapiro2000}. The basic equations for momentum, 
temperature and salinity in the upper mixed layer are similar to the integrated 
vertically equations for inner layers of the model. The commonly used convective 
adjustment scheme is applied to simulate vertical convection. According to our previous 
studies \citep{OM11} the MHI quasi-isopycnal layered model successfully 
simulates the mesoscale eddy dynamics, interaction between eddies over the shelf edge 
and steep continental slope of the Japan Basin, as well as, mesoscale eddies and currents, 
mixing and transport of water masses in the Peter the Great Bay of the Japan/East Sea \citep{FAO13}.

The numerical experiment with the MHI model is focused in this paper on simulation of the
mesoscale circulation over the deep JB, its continental slope and shelf during summer
and fall. The model domain is the closed sea area \N{39\text{--}44} and
\E{129\text{--}139} with the horizontal resolution 2.5~km \citep{OM11}.

The no-slip boundary conditions
for current velocity at the sea domain contour, including sea coast, northern, eastern
and southern boundaries, are set. The model is assumed to consist of 10 quasi-isopycnal layers
with the first one to be a horizontally inhomogeneous upper mixed layer. The first 9 layers
are located inside the main pycnocline with the lower boundary of the ninth layer to be the
lower boundary of the main pycnocline which is not deeper than 250~m in the area studied.
The lower tenth layer includes deep and bottom waters of the JES. The realistic bottom topography
(Fig.~\ref{fig1}), obtained from ETOPO2 (2-Minute Gridded Global Relief Data), is one of
the most important factor in simulation of the large scale and mesoscale circulations.

The initial conditions for summer isopycnal interfaces in the model layers,
temperature and salinity distribution include only large scale features of the
model variables. It have been taken from oceanographic survey in 1999
\citep{Talley2001,Talley2006} with a substantial smoothing. After smoothing, there were no
any mesoscale structures in the initial conditions. The MHI model has been
integrated with the time step of 4~min from the initial condition under realistic
meteorological situations. The first month with June meteorological conditions is
a time interval of the model spin up. The wind stress, short wave radiation, near
surface air temperature, humidity and wind speed have been taken from daily mean
National Centers for Environmental Prediction/National Center for Atmospheric Research 
(NCEP/NCAR) Reanalysis. The heat fluxes at the sea surface, being depended on the
mixed layer temperature and meteorological characteristics, are calculated in the model.

The coefficients of quasi-isopycnal biharmonic viscosity, harmonic viscosity,
and diffusion used in the momentum and heat/salt transfer equations have been varied
correspondingly from $10^{17}$ m$^4$s$^{-1}$, $10^{7}$ m$^2$s$^{-1}$ and
$0.4 \cdot 10^{7}$ m$^2$s$^{-1}$ in the model spin up
(60 days) to $10^{16}$ m$^4$s$^{-1}$, $10^{6}$ m$^2$s$^{-1}$ and
$0.4 \cdot 10^{6}$ m$^2$s$^{-1}$ during the other months of the warm
period of a year until late November. The quasi-isopycnal harmonic viscosity is
applied only near the domain boundary in a warm period of a year and in the whole
area in winter. We basically simulate the nonlinear large scale and mesoscale circulation over
the JB, continental slope and shelf taking into account daily mean external
atmospheric forcing from July~1 to December~1. The numerical experiments with
minimized coefficients of the horizontal and vertical viscosity show intensive
mesoscale dynamics. In particular, they demonstrate variability of the mesoscale
anticyclonic and cyclonic eddies over the shelf, the shelf break and base of the
continental slope and in the central deep area of the JB as well. The anticyclonic
eddies, generated over the shelf break and continental slope, move usually
downstream of the current with prevailing phase velocity of about 6--8 cm s$^{-1}$
\citep{OM11}. Some quasi-stationary mesoscale eddies in the central JB area,
including the eddies studied in this paper, are generated in the thermocline and deep JB
water over the mesoscale bottom troughs and sea mounts.

Figure~\ref{fig2} demonstrates the monthly mean current velocity fields in July
and October in the 6th model layer in the deep JB area (see Fig.~\ref{fig1}~a).
The system of anticyclonic and cyclonic eddies in the velocity field varies from
July to October~-- November. The most stable one is an ACE simulated in
the central region of the studied area over a mesoscale bottom trough.  Its
center is shifting with time during the model run from July to November in the
vicinity of the point with coordinates \N{41.15} and \E{134} where similar
ACEs have been often observed during the oceanographic survey in early August 1999
\citep{Talley2001,Talley2006} and in the current meter observation from August 1993 to
July 1996 at the mooring station M3 \citep{Takematsu99}. The mesoscale cyclonic eddies,
surrounding the anticyclonic one, look like short-lived eddies moving around the anticyclone
with the lifetime of about a few days.
\begin{figure*}[!htb]
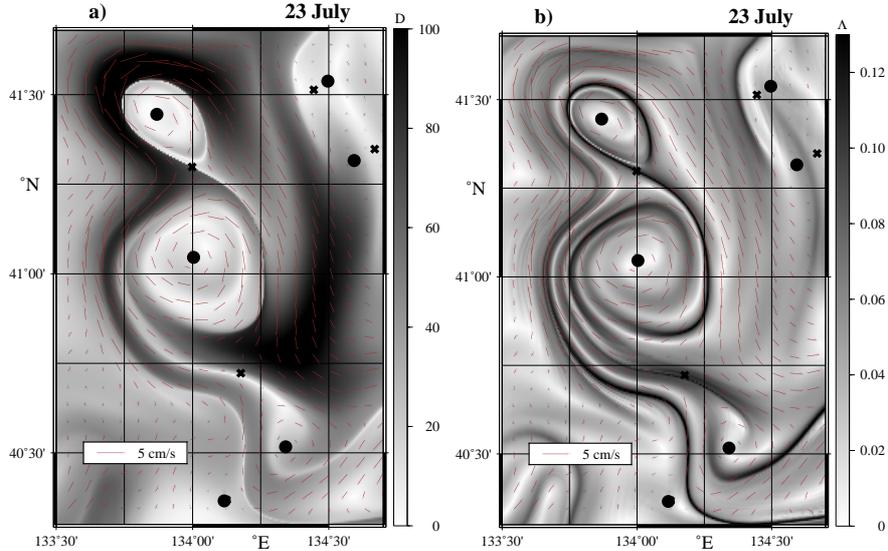

\begin{center}
\includegraphics[width=0.35\textwidth,clip]{fig3a.eps}
\includegraphics[width=0.35\textwidth,clip]{fig3b.eps}
\end{center}
\caption{Manifestations of modeled eddies on 23 July in the ninth layer
on a) the drift map ($D$ in km)
and b) the Lyapunov map ($\Lambda$ is in days$^{-1}$) with the velocity field imposed.
``Instantaneous'' elliptic and hyperbolic points, to be present in the area on 23 July,
are indicated by circles and crosses, respectively.}
\label{fig3}
\end{figure*}

\subsection{Lagrangian approach}

Two-dimensional motion of a fluid particle inside each modeled layer
is governed by advection equations
\begin{equation}
\frac{d x}{d t}= u(x,y,t),\quad \frac{d y}{d t}= v(x,y,t),
\label{adveq}
\end{equation}
where $x$ and $y$ are the longitude and latitude of the particle,
$u$ and $v$ are angular zonal and meridional components of its velocity inside a given layer.
Numerically generated velocity fields are given as discrete data sets on a grid.
So, a bicubical interpolation in space and an interpolation by third order Lagrangian polynomials
in time are used to integrate Eqs. \ref{adveq}.
The velocity components are interpolated independently on each other.
The velocities obtained are substituted in Eqs. \ref{adveq} which are integrated with a fourth-order
Runge-Kutta scheme with a fixed time step. The outputs are transformed in the geographical
coordinates to get images and maps.

A method for computing FTLE,
which is valid for $n$-dimensional vector fields, has been proposed recently
by \citep{OM11}. The Lyapunov exponents in this method are defined via singular
values, $\sigma_i(t,t_0)$, of the evolution matrix for linearized $n$-dimensional
equations of motion as follows:
\begin{equation}
\Lambda_i=\lim_{t\to\infty}\frac{\ln\sigma_i(t,t_0)}{t-t_0}, \quad i=1,2, \dots, n.
\label{lyap_def_sigma}
\end{equation}
Quantities
\begin{equation}
\Lambda_i(t,t_0)=\frac{\ln\sigma_i(t,t_0)}{t-t_0}
\label{lyap_ftle}
\end{equation}
are called FTLE which of each is the ratio of the logarithm of the maximal possible
stretching in a given direction to a finite time interval $t-t_0$.

Synoptic maps of some Lagrangian indicators have been shown to be a useful means
to visualize oceanic
features over different scales like common hydrological and
Lagrangian fronts \citep{P13,Prants2014b,Prants2014c},
strong currents \citep{Prants2014}, submesoscale  and mesoscale eddies
\citep{OM11,Prants2013}.
Such maps are obtained by integrating advection equations (\ref{adveq}) backward in time
for a given period, computing one of the Lagrangian indicators for a
large number of tracers and coding its magnitude by color. The Lagrangian indicators are
functions of particle's trajectories which carry an information about origin and history
of water masses.

The convenient Lagrangian indicator for identifying boundaries of submesoscale
and mesoscale eddies is a finite-time absolute
displacement, $D$, that is simply the distance between
the final, $(x_f,y_f)$, and
initial, $(x_0,y_0)$, positions of advected particles on the Earth sphere with the radius $R$
\begin{equation}
D\equiv R\arccos[\sin y_0 \sin y_f +\cos y_0 \cos y_f \cos (x_f - x_0)].
\label{drift}
\end{equation}
This quantity along with zonal and meridional displacements have been shown
to be useful in quantifying transport of radionuclides in the Northern Pacific
after the accident at the Fukushima nuclear power plant in March 2011 \citep{DAN11,Prants2014}
and in identifying Lagrangian fronts with favourable fishery conditions \citep{Prants2014c}.

\section{Results of simulations}
\subsection{Three-dimensional structure and evolution of
anticyclonic eddies in the Japan Basin}
\begin{figure*}[!htb]
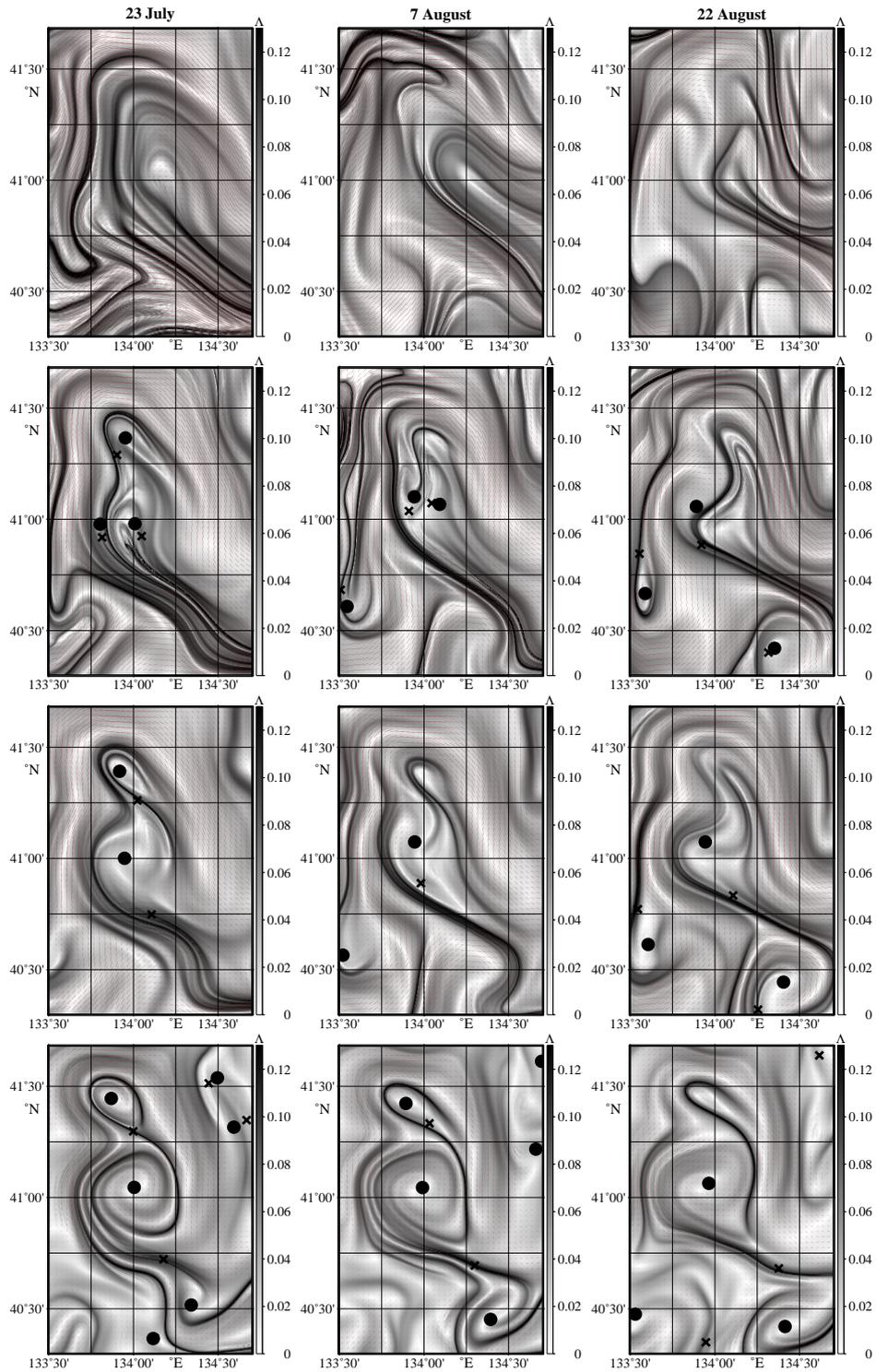

\begin{center}
\includegraphics[width=0.245\textwidth,clip]{fig4a1.eps}
\includegraphics[width=0.245\textwidth,clip]{fig4b1.eps}
\includegraphics[width=0.245\textwidth,clip]{fig4c1.eps}\\
\includegraphics[width=0.245\textwidth,clip]{fig4a3.eps}
\includegraphics[width=0.245\textwidth,clip]{fig4b3.eps}
\includegraphics[width=0.245\textwidth,clip]{fig4c3.eps}\\
\includegraphics[width=0.245\textwidth,clip]{fig4a5.eps}
\includegraphics[width=0.245\textwidth,clip]{fig4b5.eps}
\includegraphics[width=0.245\textwidth,clip]{fig4c5.eps}\\
\includegraphics[width=0.245\textwidth,clip]{fig4a9.eps}
\includegraphics[width=0.245\textwidth,clip]{fig4b9.eps}
\includegraphics[width=0.245\textwidth,clip]{fig4c9.eps}
\end{center}
\caption{Vertical eddy structure
in summer shown on 23 July, 7 and 22 August. The Lyapunov maps in the 1st, 3rd, 5th and 9th layers  are shown
from the top to the bottom, respectively.
The vortex pair, seen in the lower layers, evolves gradually with time
in a single eddy which, however, is not visible in the surface layers. The 
finite-time Lyapunov exponent $\Lambda$ is in units of days$^{-1}$. 
Elliptic and hyperbolic points are indicated by circles and crosses, respectively.}
\label{fig4}
\end{figure*}
\begin{figure*}[!htb]
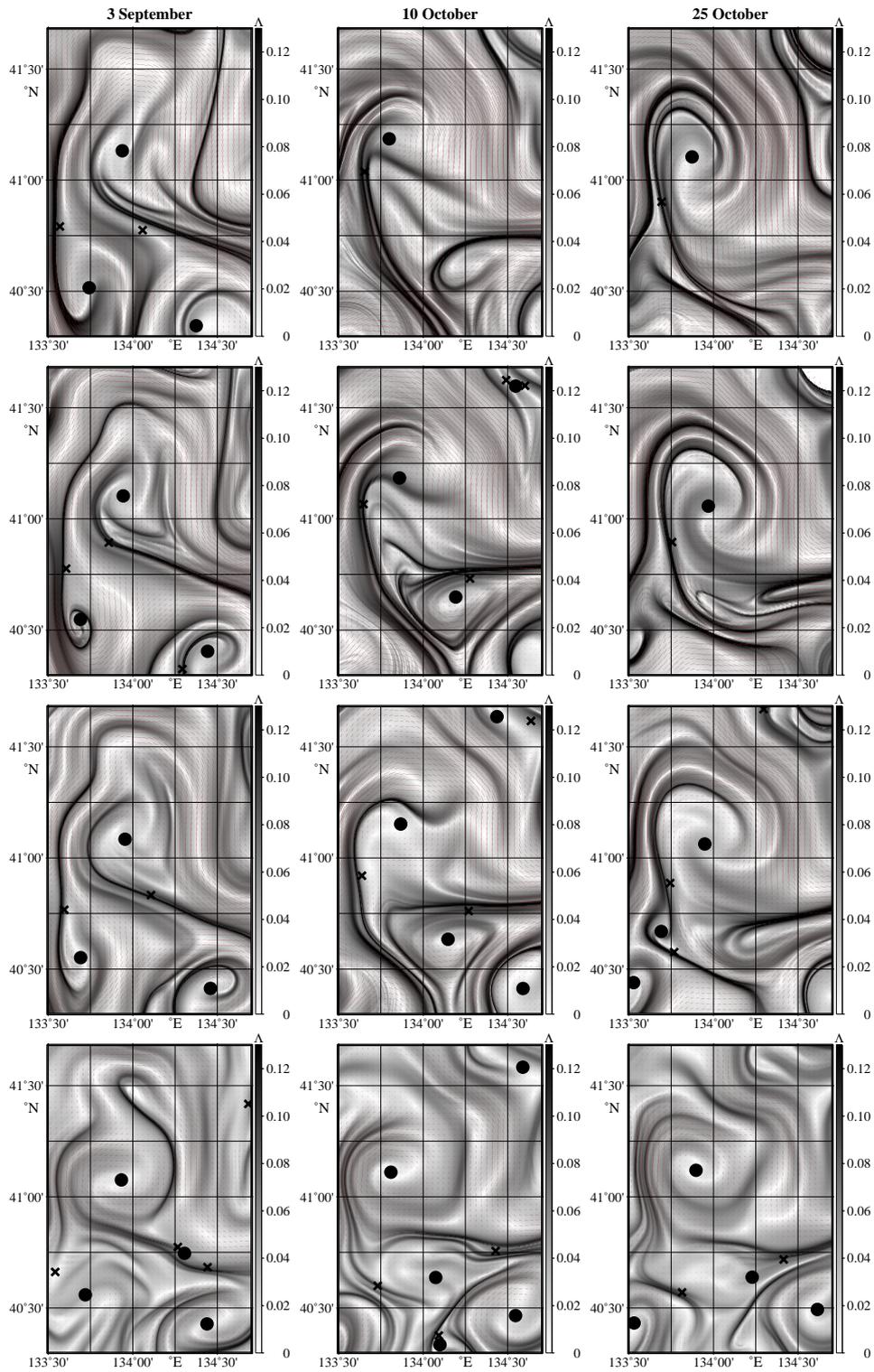

\begin{center}
\includegraphics[width=0.25\textwidth,clip]{fig5a1.eps}
\includegraphics[width=0.25\textwidth,clip]{fig5b1.eps}
\includegraphics[width=0.25\textwidth,clip]{fig5c1.eps}\\
\includegraphics[width=0.25\textwidth,clip]{fig5a3.eps}
\includegraphics[width=0.25\textwidth,clip]{fig5b3.eps}
\includegraphics[width=0.25\textwidth,clip]{fig5c3.eps}\\
\includegraphics[width=0.25\textwidth,clip]{fig5a5.eps}
\includegraphics[width=0.25\textwidth,clip]{fig5b5.eps}
\includegraphics[width=0.25\textwidth,clip]{fig5c5.eps}\\
\includegraphics[width=0.25\textwidth,clip]{fig5a9.eps}
\includegraphics[width=0.25\textwidth,clip]{fig5b9.eps}
\includegraphics[width=0.25\textwidth,clip]{fig5c9.eps}
\end{center}
\caption{The same as in Fig.~\ref{fig4} but in fall on 3 September, 10 and 25 
October. The eddy appears at the surface to the end of October.}
\label{fig5}
\end{figure*}

In this section we focus on a simulated ACE centered
at $\simeq \N{41}$ and $\simeq \E{134}$ which is visible in Fig.~\ref{fig2}
in the 6th layer velocity field averaged over July and October in a typical year of simulation.
It is a typical quasi-stationary coherent feature of the SPF, constrained by bottom topography 
(see the bathymetry map in Fig.~\ref{fig1}b), 
that has been found in this area
in satellite images \citep{Takematsu99} and measured
at a mooring station \citep{Takematsu99}
and in hydrographic surveys \citep{Talley2001,Talley2006}.

Eddy identification in our study is based on geometry of the velocity field
and specific features of the FTLE and displacement fields.
The following necessary and sufficient conditions we use to identify an eddy center
and its boundaries. The velocity magnitude has zero value at the eddy center which is defined as
an elliptic stagnation point whose stability is checked daily by the standard methods.
The zonal and meridional rotational velocities have to
reverse in sign across the eddy center, and their magnitudes have to increase initially
away from it, go through a maximum and then decrease.
Eddy boundaries are identified better in the FTLE and displacement fields
than in the corresponding velocity field.


Eddy core and surrounding water masses may have a rather distinct origin that can be traced out
with the help of so-called drift maps introduced by \citet{DAN11,P13,Prants2014}. The map for absolute
displacements of tracers is computed as follows. A region under study is seeded with
a large number of synthetic tracers for each of which advection equations are integrated backward in time
for an appropriate period, 30 days in our case. The distance, $D$, between
positions of each tracer on the first and last days of that period is calculated
in each model layer
as (\ref{drift}) and coded by color.
Such a map in Fig.~\ref{fig3}~a clearly demonstrates a vortex pair with two
ACEs in the 9th layer with elliptic points at their centers
(\N{41}, \E{134}) and (\N{41.4}, \E{133.9}) and a hyperbolic
point in between (\N{41.3}, \E{134}).
The boundary of the northern eddy core can be identified
as a closed curve with the maximal local gradient of $D$ that separates waters,
involved in the rotational motion around the vortex center, from ambient waters.
Magnitudes of the absolute
displacements for the former ones are much smaller than for the particles outside the eddy.

The southern eddy has a more complicated structure because of its intensive interaction
with ambient waters during the month of integration.
The drift maps allow to delineate the transport corridors by which eddies
can gain water from their surroundings. They look like dark ``tongues'' in Fig.~\ref{fig3}a
encircling that eddy.
The origin of those waters can be traced out by computing tracer's displacements in
zonal and meridional directions backward in time for the month, from 23 July to 23 June.
The corresponding zonal and meridional
drift maps shown in Fig.~S1 allow to visualize where the waters in the ninth layer
came from. Blue color of the water ``tongues'' around that eddy mean that it
trapped waters from the south (Fig.~S1a) and east (Fig.~S1b).
Complementary backward-in-time Lagrangian longitudinal (Fig.~S2a) and latitudinal (Fig.~S2b) drift maps
show by color the longitudes and latitudes, respectively, from which tracers,
initialized in the area on 23 June, came to their final positions on 23 July.

The sharp boundary between waters with high gradients of
a Lagrangian indicator (e.~g., the absolute displacement $D$) was
called a ``Lagrangian front'' \citep{P13,Prants2014b}.
The Lagrangian fronts, encircling each of the eddies in the pair in Fig.~\ref{fig3}a,
can be identified by a narrow white stripe demarcating
the curve with the maximal gradient of $D$. White color
means that the corresponding particles have experienced very small displacements
over the period of integration because they rotated around eddy's centers.
The sizes of the southern and northern eddies are estimated to be $\approx 35\times 45$~km
and $\approx 20\times 20$~km, respectively.


The FTLE field provides complementary information on the horizontal eddy structure.
We computed it in the same area by the method \citep{OM11} and coded the values of
the FTLE, $\Lambda$, by color.
The map in Fig.~\ref{fig3}b demonstrates the same vortex pair surrounded by black ``ridges''
with local maximal values of $\Lambda$ which are known to approximate
unstable manifolds of the hyperbolic trajectories in the region \citep{HY00}.
There are typically a number of
hyperbolic trajectories around eddies in the ocean which are connected with the corresponding
hyperbolic stagnation points marked by crosses on our Lagrangian maps. Each of which has its
own unstable manifold that is manifested on backward-in-time FTLE maps as a black ``ridge''
\citep[for a review on chaotic advection in the ocean see][]{Wiggins05,KP06}.

In oceanic flows trajectories and stagnation points can gain or lose hyperbolicity
over time.
It means, in particular, that hyperbolic stagnation points may appear and disappear
in the course of time.
Only those ones, which exist on 23 July, are shown in Fig.~\ref{fig3}. As to the southern eddy,
it is confined from the east and south by the S-like unstable manifold of the hyperbolic
point located between the eddies in the vortex pair. Any unstable manifold influences strongly on
adjacent fluid parcels. It is illustrated in Fig.~S3, where we placed blue and rose patches with
tracers near the S-like unstable manifold, and advected them forward in
time for three and half months. Both the patches elongate along that manifold.
The red patch was chosen in the center of the northern eddy and is shown in Fig.~S3
to deform slightly in the course of time until the northern eddy begins to break down
to the end of summer. The southern eddy is confined from the west by
the unstable manifold of a hyperbolic point which disappeared to 23 July but existed before.
The complicated pattern of the black ``ridges'' around the southern eddy in Fig.~\ref{fig3}b
confirms the conclusion about its intensive interaction with ambient waters.


To illustrate the vertical structure of the vortex pair and its evolution
we show in Fig.~\ref{fig4} the summer FTLE maps in the first, third,
fifth and ninth layers. On 23 July the pair with two
ACEs, elliptic points at their centers and a hyperbolic
point in between is clearly seen in the lower layers below the 4th one.
The vortex pair is especially prominent in the 9th layer that is a lower
boundary of the main pycnocline. It is not
recognizable in the third layer  and above where the corresponding
hyperbolic point between the eddies disappears, but the elliptic points still exist at their places.
The elliptic points are absent at the surface where there are no signs of vortex motion.

In the course of time the pattern changes. The pair gradually decays
in the sense that the northern eddy merges with the southern one (compare the map on 7 August with
the map on 22 August when the northern elliptic point disappears in the 9th layer).
As to the other layers, it is seen that
the northern elliptic point disappears earlier. It is possible to recognize to 22 August
a single ACE with the size $\approx 40 \times 50$~km in the 5th layer and below.
Thus, we have got to the end of summer the ACE not reaching the surface.

Changes in the vertical vortex structure in September and October are shown in
Fig.~\ref{fig5}. The eddy in fall is clearly visible in the 5th layer and below as a single ACE
of the same size as in August. As to the surface layers, a prominent eddy structure becomes visible
there only to the end of October.
In the beginning of September the elliptic point appears in the first layer
at the place where the eddy is visible in the 5th layer and below, but the eddy cannot
be clearly detectable on the corresponding Lyapunov maps.
Thus, the bowl-shaped eddy is formed to the end of October. It penetrates from the surface to the
bottom gradually decaying to the end of November.

\subsection{Zonal and meridional layer interface and temperature vertical sections
across the simulated eddy}
\begin{figure*}[!htb]
\begin{center}
\includegraphics[width=0.253\textwidth,clip]{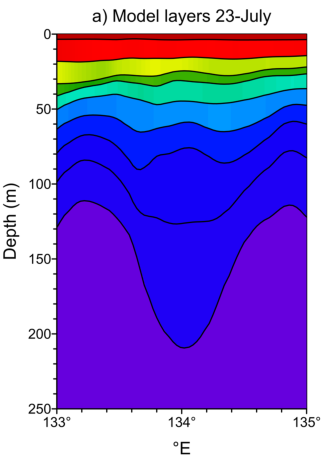}
\includegraphics[width=0.253\textwidth,clip]{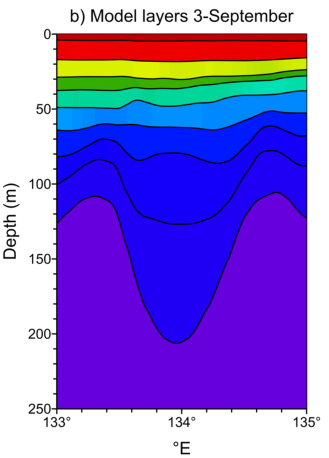}
\includegraphics[width=0.253\textwidth,clip]{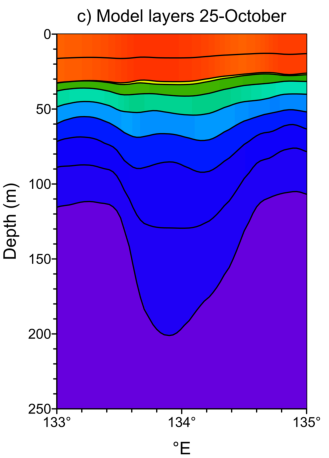}\\
\includegraphics[width=0.253\textwidth,clip]{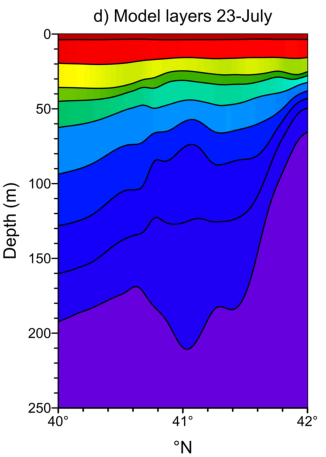}
\includegraphics[width=0.253\textwidth,clip]{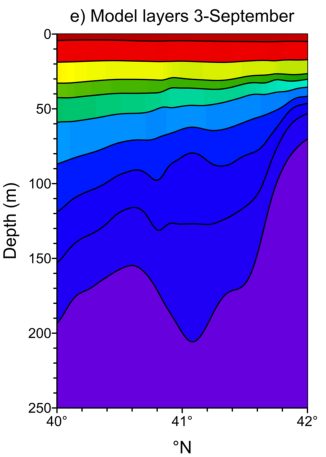}
\includegraphics[width=0.253\textwidth,clip]{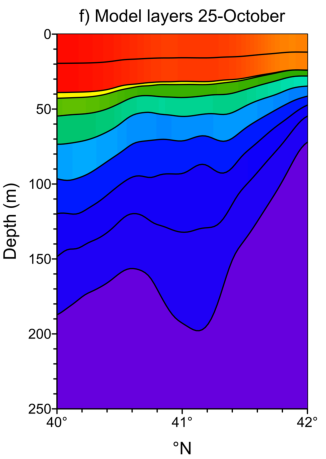}\\
\caption{Zonal (along \N{41}, the upper row) and
meridional  (along \E{134}, the lower row) sections of the interfaces between modeled eddy layers
on 23 July, 3 September and 25 October. Each quasi-isopycnal layer is shown by 
its own color.}
\label{fig6}
\end{center}
\end{figure*}
\begin{figure*}[!htb]
\begin{center}
\includegraphics[width=0.253\textwidth,clip]{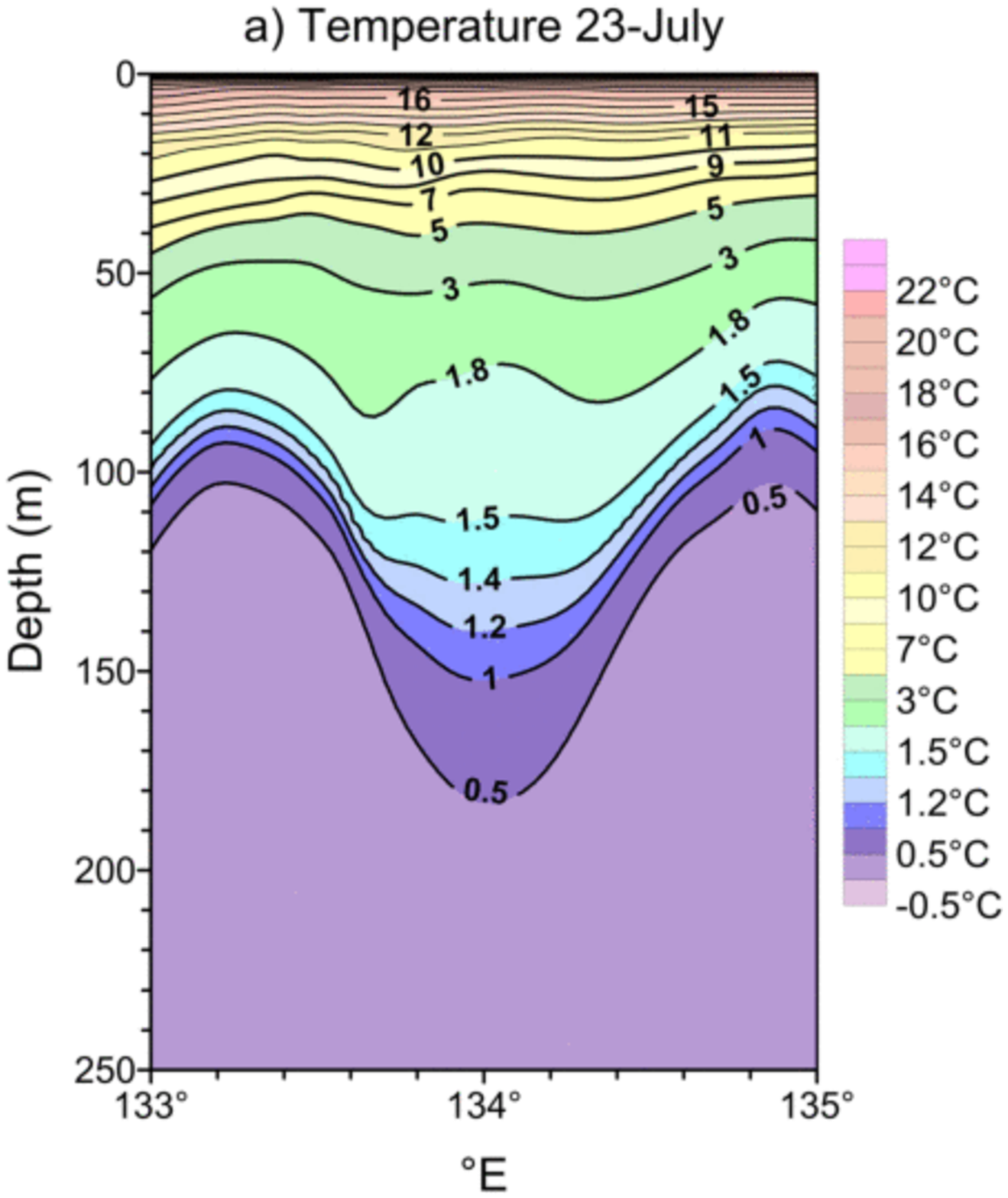}
\includegraphics[width=0.253\textwidth,clip]{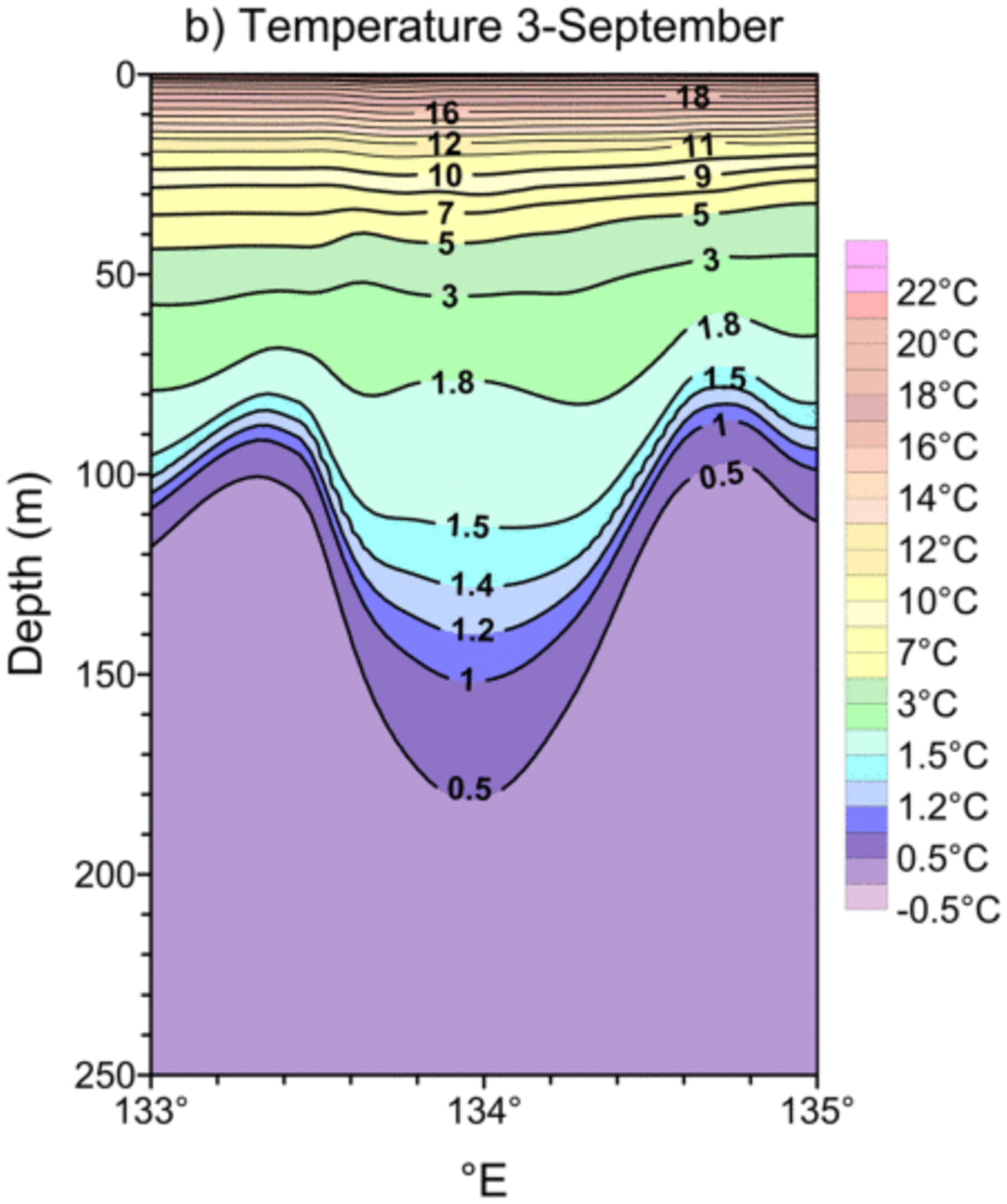}
\includegraphics[width=0.253\textwidth,clip]{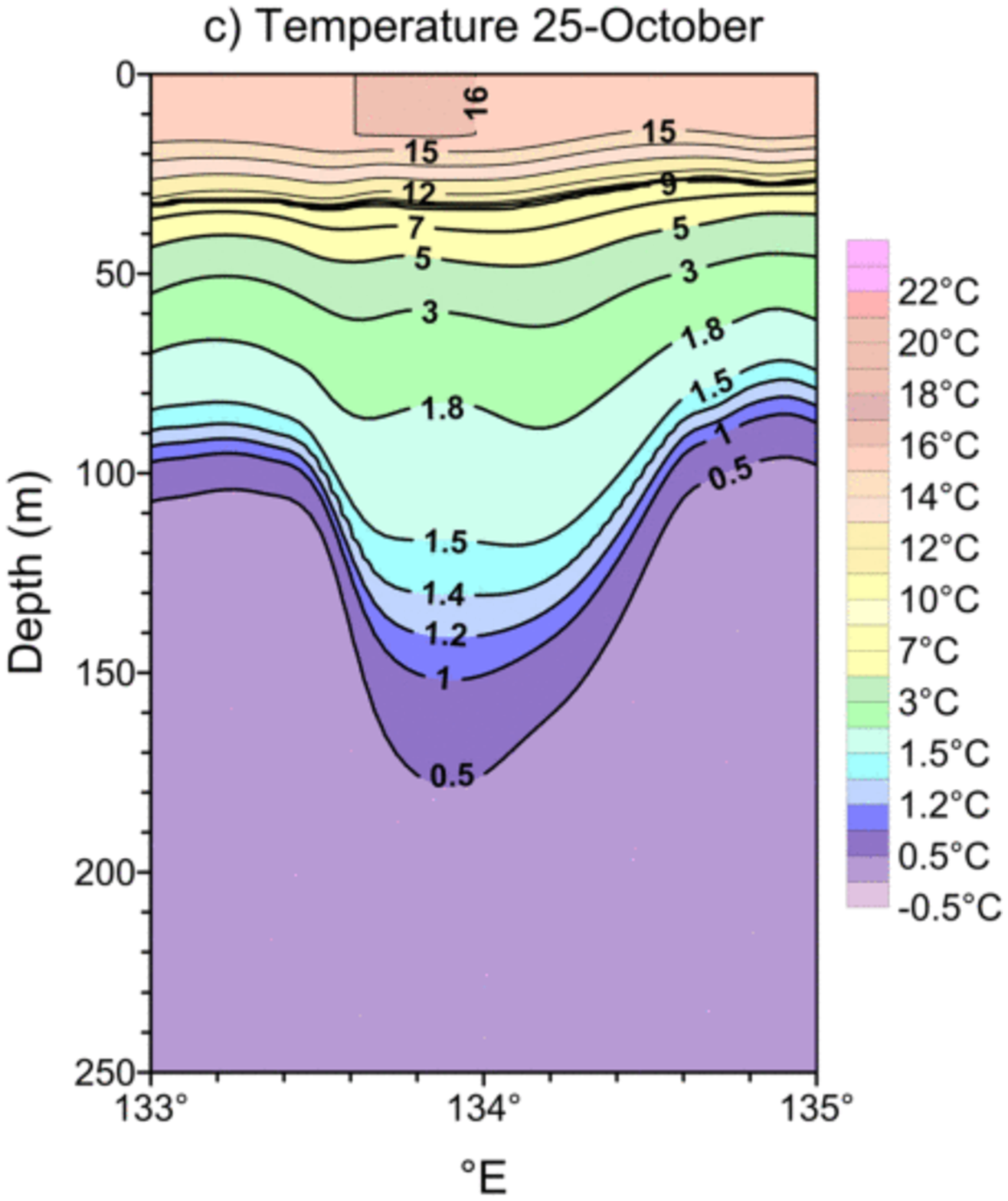}\\
\includegraphics[width=0.253\textwidth,clip]{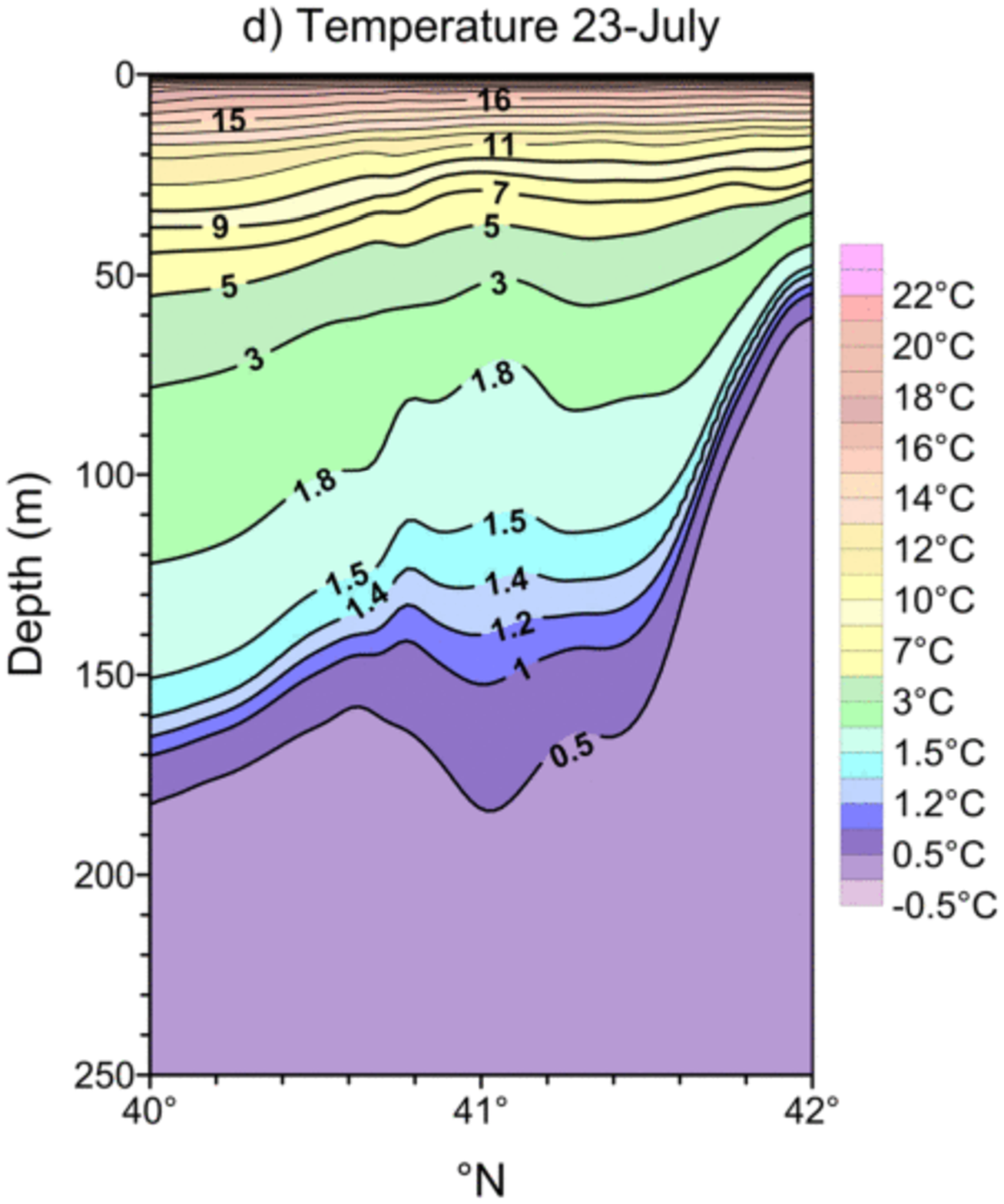}
\includegraphics[width=0.253\textwidth,clip]{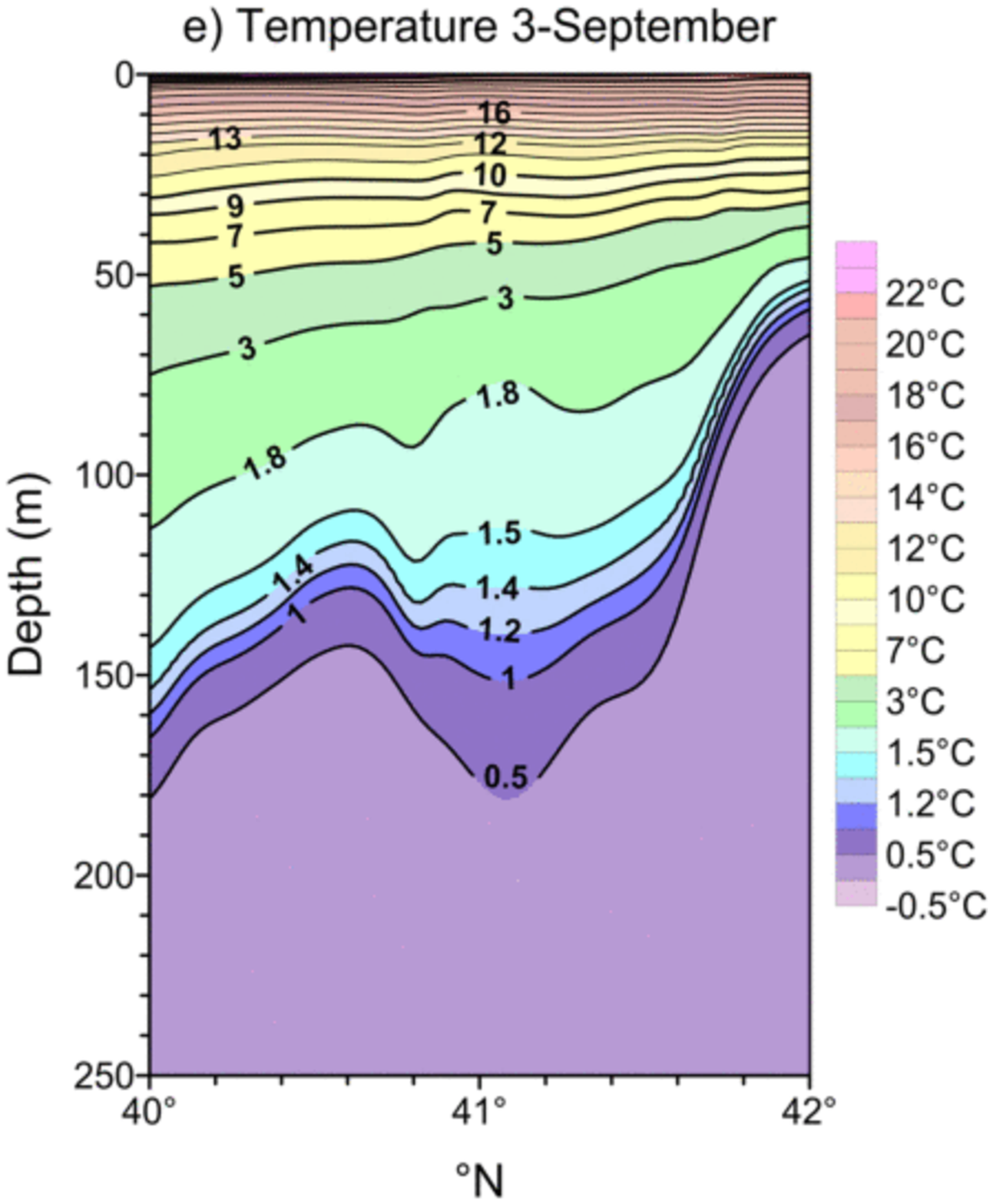}
\includegraphics[width=0.253\textwidth,clip]{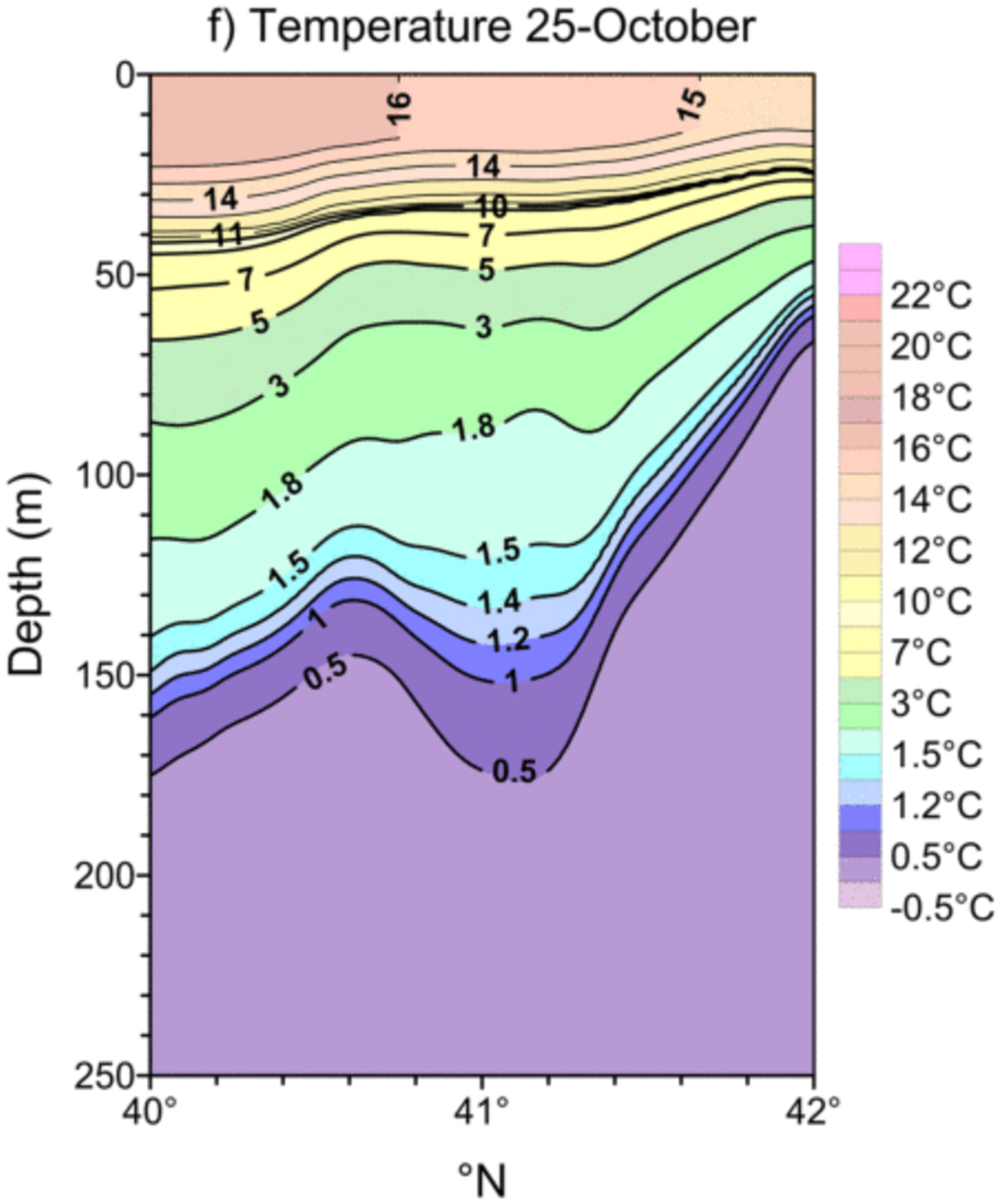}\\
\caption{The same as in Fig.~\ref{fig6} but for temperature
sections across the simulated eddy on 23 July, 3 September and 25 October.}
\label{fig7}
\end{center}
\end{figure*}
\begin{figure*}[!htb]

\begin{center}
\includegraphics[width=0.9\textwidth,clip]{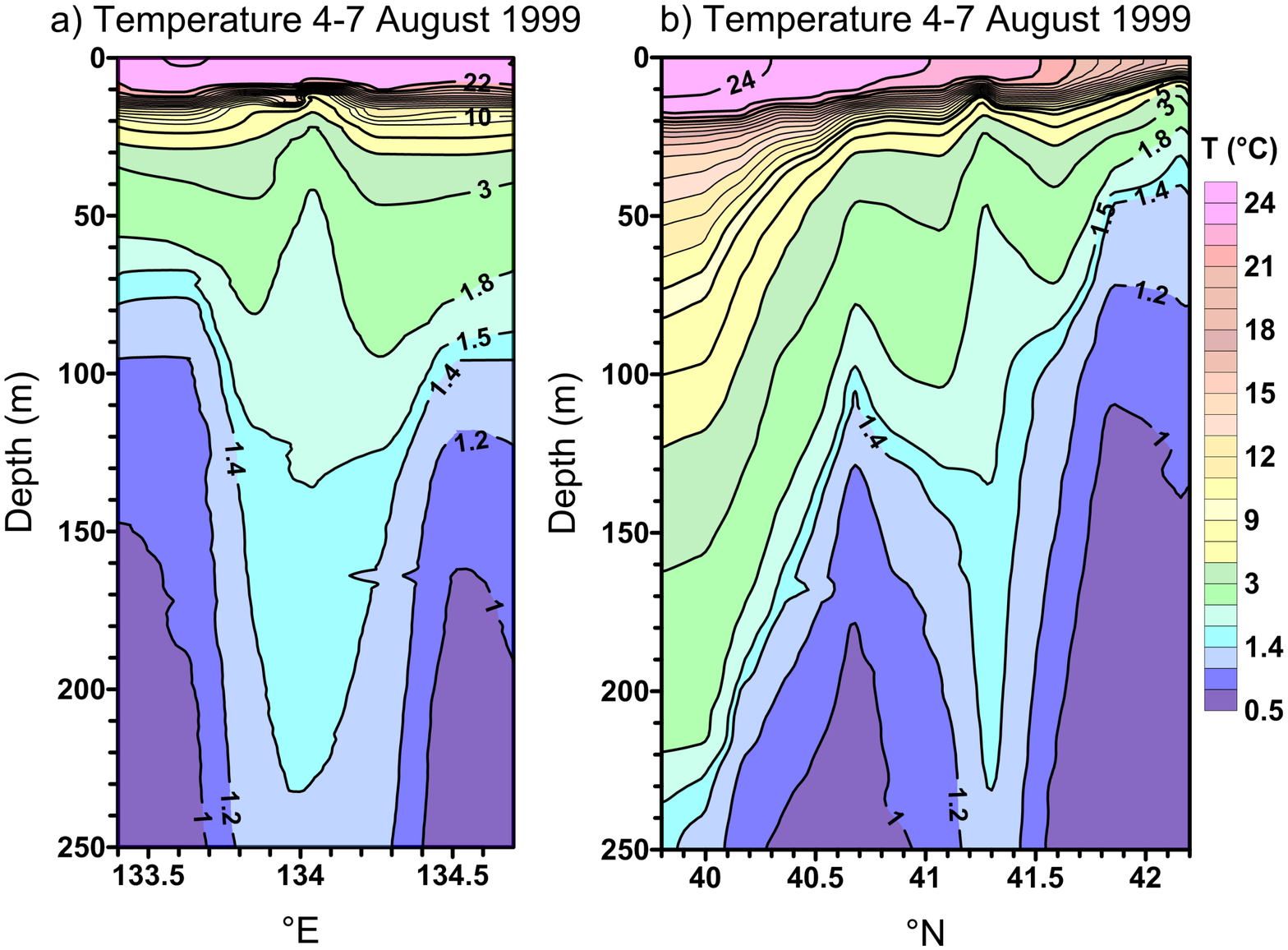}
\end{center}
\caption{a) Zonal (along \N{41}) and b) meridional
(along \E{134}) observed sections of temperature across
the anticyclonic eddy to be found
during the oceanographic CTD-hydrochemical survey in early August 1999 \citep{Talley2001}.}
\label{fig8}
\end{figure*}

Ten layers have been used in the MHI model adopted to the JES.
In order to illustrate transformation of our vortex structure with time,
we compute vertical zonal and meridional sections across the simulated eddy.
Figures~\ref{fig6} and \ref{fig7} show evolution of its vertical structure
from late July to late October in terms of sections of quasi-isopycnal layer interfaces
(Fig.~\ref{fig6}) and water temperature in the layers (Fig.~\ref{fig7}). The anticyclonic eddy,
illustrated on the Lagrangian maps in Figs.~\ref{fig3}--\ref{fig5}, is formed over the
mesoscale bottom trough in early July at first in the main pycnocline and underlying layers
and presents within the layers from the 4th one and below all the time after its formation.

The Lagrangian maps on 23 July in Fig.~\ref{fig3} clearly show the vortex
pair oriented in the meridional direction. The zonal sections in Fig.~\ref{fig6}
along the latitude \N{41} cross the southern ACE which is manifested
as a depression of the layers from the 3rd to 9th ones around the elliptic point.
The vortex pair evolves to
3 September to a single ACE (see Fig.~\ref{fig5}) whose zonal cross section is shown
in Fig.~\ref{fig6}b. It still does not extend to the surface.
To 25 October the eddy reaches the surface (see Figs.~\ref{fig5} and
~\ref{fig6}c). The northern ACE of the vortex pair, seen in Fig.~\ref{fig3} on 23 July,
is hardly visible in the meridional cross section as a small depression in the lower layers
to the right of the main depression (Figs.~\ref{fig6}d and e). 

From July to early September, the ACE presents only occasionally in the numerical solution in
the upper mixed layer (layer~1) and in the seasonal pycnocline (layers~2--4).
In this period of warm season, when the upper mixed layer is comparatively
thin (see Figs.~\ref{fig6}a, b, d and e)
and the seasonal pycnocline is very strong (see Figs.~\ref{fig7}a, b, d and e),
the simulated ACE is unstable in the
upper layers. Its lifetime in these layers does not exceed a few days. During summer, it
episodically appears and breaks down into smaller submesoscale eddies in the upper
layers. During October~-- November, when the thickness of the upper mixed layer
increases from 10 to 15--30~m (Figs.~\ref{fig6}c and f)
and the seasonal pycnocline is weak, the ACE becomes as stable in the upper layers
as in the underlying ones. It is also manifested in the zonal temperature section in
Figs.~\ref{fig7}c as the closed isotherm at the sea surface.

\subsection{Zonal and meridional temperature vertical sections across
the observed eddy}

During the oceanographic CTD-hydrochemical survey
in summer 1999 \citep{Talley2001}, the mesoscale ACE (with the center approximately
at the same place as the M3 mooring \citep{Takematsu99} and our simulated eddy) has been observed in
temperature, salinity, dissolved oxygen and NO${}_3$
sections along \E{134} and \N{41.25}.
The warm fresh core of the eddy with high gradients in temperature, salinity and dissolved
oxygen at its edge was situated in the thermocline within the layer from 50 to 150~m.

Figure \ref{fig8} shows water temperature structure of that ACE
in zonal and meridional sections of the oceanographic survey
in early August 1999 \citep{Talley2001}. The density gradient in this
eddy basically depends on the temperature gradient. The eddy core, to be surrounded
by maximal temperature and density gradients, looks like a lens.
The eddy occupies the water column below the seasonal pycnocline (30--50~m).
The anticyclonic eddy was not clearly observed in the upper mixed layer
and in the seasonal pycnocline both in the zonal and meridional temperature sections.
It is an important feature of the observed eddy closely related to the simulation
results discussed above.
Eddies with similar vertical structure in temperature and density cross-sections,
named as intra-thermocline eddies, have been observed south of the Subarctic Front
over the western side of the Yamato Rise and within quasi-stationary meanders of
the Tsushima Current \citep{Gordon2002}. They have been successfully
simulated in this area by \citep{Hogan2006} using an isopycnal ocean circulation model.
The observed eddy in Fig.~\ref{fig8} is situated over the mesoscale bottom trough
surrounding by sea mounts in the western area of the JB (see Figs.~\ref{fig1}a and
b) practically in the same area as our simulated eddy.
In summer, the position and the vertical structure of the simulated ACE in Figs.~\ref{fig6} and \ref{fig7}
are similar to those for the observed ACE in Fig.~\ref{fig8}. Both the simulated and
observed ACEs have the similar eddy core, the relief of layer interfaces and isotherms.

The observed ACE is stronger than the simulated one with a much more
strong temperature/density front situated to the south of the ACE
and a stronger vertical stratification. That difference may be explained by
the fact that the observation has been made
in the warm climatic period and warm year (1999), whereas the simulation
has been performed under daily meteorological conditions averaged
over 25-years period from 1976 to 2000. Moreover, we did not take into account
meridional heat advection from the southern sea area and the southern boundary
of the model domain due to no-slip boundary condition for the current velocity.

\subsection{Tracking maps}
\begin{figure*}[!htb]
\begin{center}
\includegraphics[width=0.35\textwidth,clip]{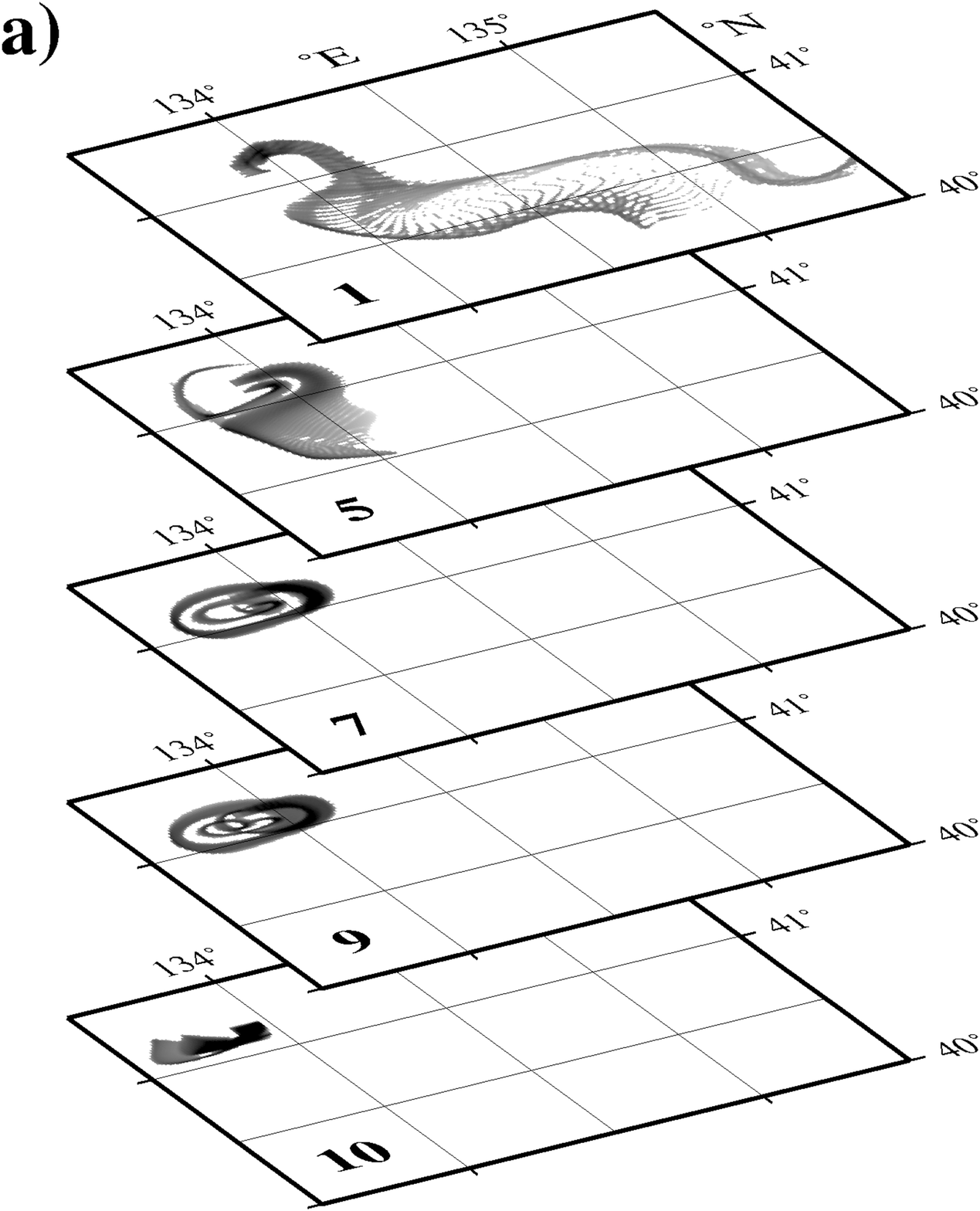}
\includegraphics[width=0.35\textwidth,clip]{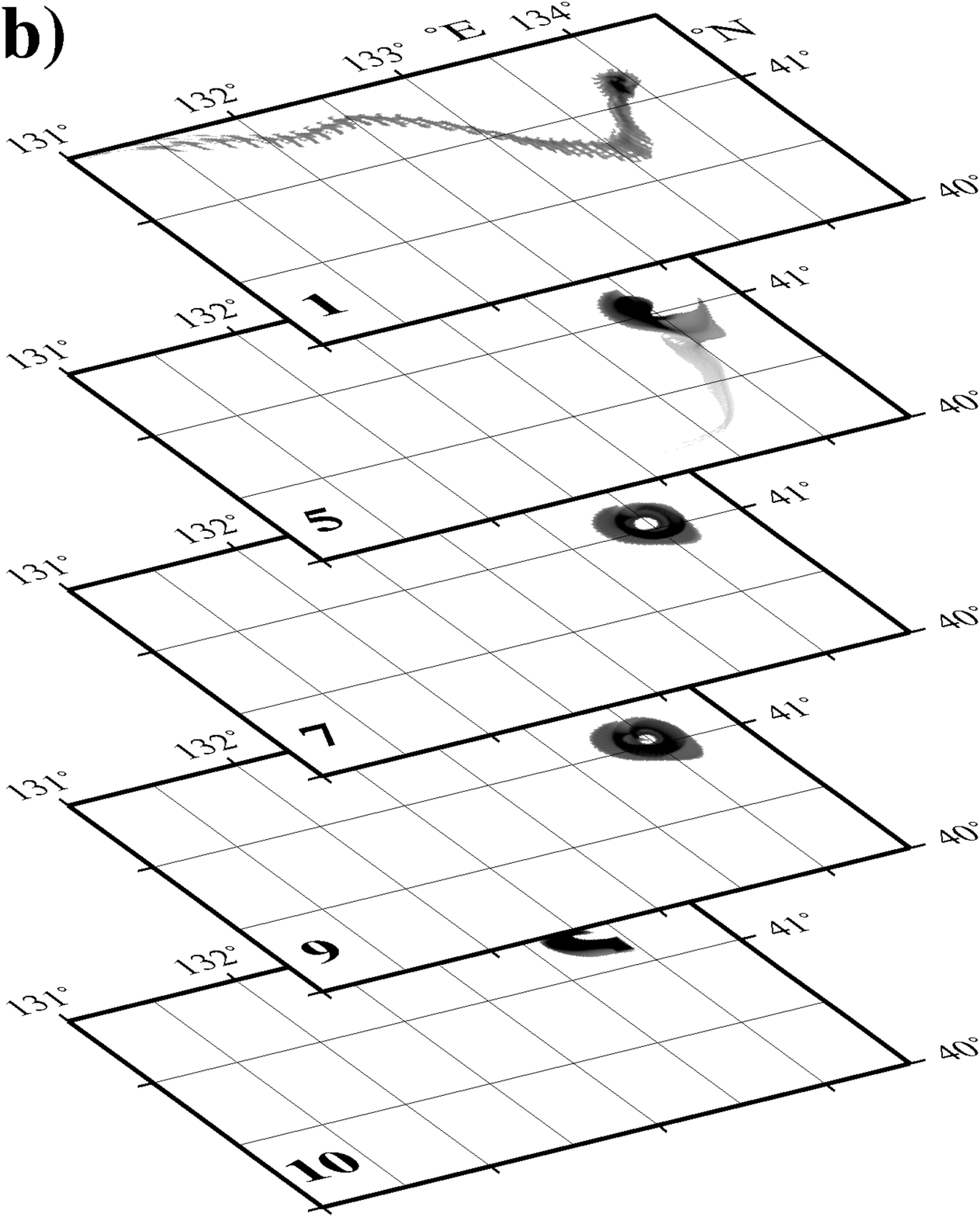}
\end{center}
\caption{Tracking maps for the markers distributed over the eddy's core on 1 September in 1st, 5th, 7th, 9th
and 10th layers numbered in the left lower corners. a) The forward-in-time map shows where
the corresponding markers were walking from 1 September to 1 November. b) The backward-in-time map shows
where they were  walking from 1 September to 1 July.
The density of traces is in the logarithmic scale.}
\label{fig9}
\end{figure*}

In this section we apply a special Lagrangian technique \citep{FAO13,Prants2014}
to visualize the origin and fate of water masses
in the eddy core. A large number of markers (250\,000) is distributed on 1 September in each layer
around the eddy center inside the patch $7\times 11$~km (\E{133.87\text{--}133.97},
\N{41\text{--}41.1}). They are advected for two months backward and forward in time
by the velocity field in each layer. The tracking maps are computed as follows.
The region under study, \E{131\text{--}136} and \N{40\text{--}43.5} is
divided in $500 \times 500$ cells. One fixes each day which cells and how many times
the markers have visited from 1 September to 1 July (backward-in-time tracking maps)
and from 1 September to 1 November (forward-in-time tracking maps).

The results are shown in Fig.~\ref{fig9} and may be interpreted as follows.
The forward-in-time tracking maps in Fig.~\ref{fig9}a show where the corresponding markers in each chosen layer
were walking from 1 September to 1 November. In this period,
the markers from the patches in the lower layers, including the 5th one,
rotated around the eddy center
with an insignificant flow outside. The ``tails'' at the upper levels mean that
at the surface the eddy expelled water from its periphery in different directions,
however, preserving its core.

The backward-in-time tracking maps in Fig.~\ref{fig9}b
show where the corresponding markers in each chosen layer
were walking from 1 September to 1 July. The patches with markers in the lower layers,
from the 7th to 10th,
practically did not change their form for two months in the past. It means that
the eddy existed in those layers
all that time at the same place without exchanging by the core water with its surroundings.
The ``tail'' of the patch in the 5th layer means that the eddy in this layer gained the water
from the south,
but its core have been at the same place for the two months (see the FTLE maps
in the 5th layer in Figs.~\ref{fig4} and \ref{fig5}). As to the surface layer  (the upper plane
in Fig.~\ref{fig9}b), markers from the initial patch in the 1st layer came to their positions on
1 September from the west.

\subsection{The form, nonlinearity and stability of simulated anticyclones}

The ACE under study is an eddy-like feature in a region of the JB 
to the north of the Yamato Rise (see Fig.~\ref{fig1}~a). That eddy has been found to be 
practically stationary for a half-year integration period including summer and fall. 
It is seen from the lower panels in Figs.~\ref{fig4} and \ref{fig5} that displacement 
of its elliptic point for three months did not exceed 10~km. 
Inspection of the AVISO velocity field at the sea surface 
for 1992 -- 2014 (http://www.aviso.oceanobs.com) 
has shown that surface eddy-like anticyclonic features often appeared in the area 
around ($\approx$ \E{134}, $\approx$ \N{41}) in cold seasons and disappeared in warm ones. 
No significant directed transport of such eddies has been found in those altimetric data.  
Taking into account these findings, complicted bottom topography in the area with 
underwater seamounts and trenches (see Fig.~\ref{fig1}~b) and observations by \citep{Takematsu99}, 
it is reasonably to suppose that we deal with the ACE constrained by bottom topography.

Computation of the FTLE and particle's displacement in each depth layer 
clearly shows that the simulated ACE evolves of the eddy, that does not reach the surface
in summer, into a one reaching the surface in fall. The corresponding elliptic points,  
demarcating the eddy's center, are absent in upper layers in summer (Fig.~\ref{fig4}) 
and appear in fall (Fig.~\ref{fig5}). This finding is confirmed by computing deformation 
of the model layers and the temperature cross sections. Whether the eddy reaches the surface 
or not, it depends on the stratification measure in the thermocline, topographic and other parameters.
In summer, when the upper mixed layer is comparatively thin and the stratification of seasonal 
pycnocline is very strong, 
the simulated eddy is unstable in the upper layers.
In fall, when the stratification of seasonal pycnocline is much weaker than in summer, the eddy becomes as stable in the upper layers
as in the underlying ones. 

The eddy must be sufficiently nonlinear to exist as a stable entity.
The measure of the nonlinearity is the so-called quasigeostrophic
nonlinearity parameter $Q_{\beta}$,
which is the ratio of the relative vorticity advection to the planetary vorticity advection
\citep{Chelton2011} defined as $Q_{\beta}=U/\beta L^2$, where $U$ is maximum rotational
speed, $L$ the eddy radius and $\beta = df/dy$ is the planetary vorticity gradient.
Let us estimate the quasigeostrophic
nonlinearity parameter of our simulated eddy in the JB. Taking
$U=0.05\text{--}0.1$~m~s$^{-1}$, $L=30000$ m, $R^2 = 9 \cdot 10^8$ m$^2$, and $\beta
= 1.73 \cdot 10^{-11}$~m~s$^{-1}$, one gets  $Q_{\beta}=3.3\text{--}6.6$.
This range of values means that the relative vorticity dominates and
suggests that our simulated eddy is nonlinear and may persist as a coherent structure.

\section{Conclusions}

We numerically investigated the vertical structure of
simulated mesoscale anticyclonic eddies often observed \citep{Takematsu99,Talley2001,Talley2006}
to the north of the Subpolar Front in the Japan/East Sea.
The model used for experiments was the MHI hydrothermodynamic quasi-isopycnal, eddy-resolved model of
a multilayer ocean \citep{Mikhailova1993,Shapiro2000} with the horizontal resolution of 2.5~km.
The Lagrangian approach has been applied to study temporal evolution and a vertical structure
of the anticyclonic vortex feature constrained by the bottom topography in the western area 
of the deep Japan Basin.
The finite-time Lyapunov exponent field along with the field of tracer's displacements
have been shown to demarcate Lagrangian boundaries of the eddy, pathways and barriers organizing
transport and mixing processes at the mesoscales and submesoscales.

Estimating the quasigeostrophic
nonlinearity parameter of our modeled eddy, we have shown that
it was sufficiently nonlinear to exist as a stable entity.
The quasi-3D analysis has been performed in the period from July to November in a model 
period of a year
when the eddy, visible in summer in the lower layers only, gradually changed its
vertical structure to become visible in the surface layers in fall.
In summer, when the upper mixed layer is comparatively thin and the seasonal pycnocline
is very strong, the simulated eddy was shown to be unstable in the upper layers.
During October~-- November, when the thickness of the upper mixed layer
increases and the seasonal pycnocline is weak, that eddy becomes as stable in the upper layers
as in the underlying ones. We applied a special
tracking technique to find out transport pathways by which the eddy exchanged
water with its surrounding at each depth level. It was found that at the surface the eddy
gained and expelled water much more intensively than in the deep layers.

The results  of simulation were compared with observed temperature zonal and meridional
cross sections of a real anticyclonic eddy to be studied at that place
during the oceanographic CTD-hydrochemical survey in summer 1999
\citep{Talley2001,Talley2006}. The position and the vertical structure of
the simulated ACE have been found to be similar to those for the observed one.
Both the simulated and observed eddies have the similar eddy core, the relief of
layer interfaces and isotherms.
The results of this paper would thus seem to indicate that topographically 
controlled eddies reaching the surface may be a state to which submerged eddies might evolve  
given sufficiently time.

\section*{Acknowledgments}

This work was supported by the Russian Foundation
for Basic Research (project nos.~12--05--00452a, 13--05--00099a and 13--01--12404ofim) 
and by the Prezidium of the Far-Eastern Branch of the Russian Academy od Sciences
(project nos.~12-I-P23-05 and 12-III-A-07-129).

\section*{Appendix A}

Supplementary materials associated with this paper can be found in the online version.

\end{document}